\documentclass[prb,twocolumn,showkeys,amsmath,amssymb]{revtex4}
\usepackage{graphicx}
\usepackage{dcolumn}
\usepackage{bm}

\begin{document}
\title{Cyclotron resonance of correlated electrons in semiconductor heterostructures} 
\author{Dinh Van An} 
\email{divan@cmp.sanken.osaka-u.ac.jp}
\altaffiliation{Present address: Department of Condensed Matter Physics, Institute of Scientific and Industrial Research, Osaka University, Mihogaoka 8-1, Ibaraki, Osaka 567-0047, Japan} 
\author{Motohiko Saitoh}
\affiliation{%
Department of Physics,~~Graduate School of Science,~~Osaka University,\\ 
1-1 Machikaneyama, Toyonaka 560-0043, Japan.
}%




\begin{abstract}%
The cyclotron resonance absorption of two-dimensional electrons in semiconductor heterostructures in high magnetic fields is investigated. It is assumed that the ionized impurity potential is a dominant scattering mechanism, and the theory explicitly takes the Coulomb correlation effect into account through the Wigner phonons.  The cyclotron resonance linewidth is in quantitative agreement with the experiment in the Wigner crystal regime at $T=4.2$K. Similar to the cyclotron resonance theory of the charge density waves pinned by short-range impurities, the present results for the long-range scattering also show the doubling of the resonance peaks. However, unlike the case of the charge density waves, our theory gives the pinning mode independent of the bulk compressibility of the substrate materials.
\end{abstract}

\keywords{%
Cyclotron resonance; Semiconductor heterostructure; Quantum transport; Wigner crystal
}

\maketitle

\section{\label{sec:intro}Introduction}

Two-dimensional (2D) electron system in strong magnetic fields has been studied intensively for many years \cite{Chui94,Tsui99}.  There are  numerous experimetal and theoretical investigations concentrating on the low density electron system in high magnetic fields when the electron correlation becomes strong and the system can be considered as a Coulomb liquid or Wigner lattice\cite{Dykman82,Saitoh87,Wilen88,Chou88,Li97,Teske99,Monarkha2000}. In order to study the characteristics of the 2D electrons in a magnetic field, the cyclotron resonance (CR) is considered as one of the basic techniques, and important information about the 2D electron system can be obtained from the CR. Though there are many theoretical works  considering the CR in the highly correlated electron liquid and correlated electrons which form a Wigner crystal on the surface of liquid helium \cite{Dykman82,Saitoh87,Teske99, Monarkha2000}, the CR of highly correlated electrons in semiconductor heterostructures is still an open question.

 In a magnetic field normal to the plane, the kinetic energy of 2D electrons is  quenched and electrons execute a cyclotron motion with orbit radius propotional to the magnetic length $l_c=\sqrt{\hbar c/eB}$ which decreases with increasing magnetic field $B$. Once $B$ becomes sufficiently high so that $l_c$ is small compared with the typical mutual distance between electrons, the crystallization of the electron system into a Wigner crystal is expected if the electron density is sufficiently low. In semiconductor heterostructures, when the mutual Coulomb energies are dominant, a Wigner crystal or a charge density wave may be formed in the quantizing magnetic field limit. Recently, an insulating state was observed on monolayer systems \cite{Willett89,Goldman90,William91}, such a state is interpreted as a Wigner crystal pinned by the disorder in the sample \cite{Chitra2001}. 
 
 The early theory by Fukuyama and Lee\cite{FL78}, which considered the pinning of the charge density waves of the 2D electrons in heterostructures under the influence of the magnetic field, has predicted that the absorption lineshape of the system has double resonance peaks: one is located at the cyclotron frequency and the other related to the impurity pinning. In this theory, the magnetic field is treated classically and the impurity potential is assumed as short-ranged ones, and the pinning mode depends on the bulk compressibility of the substrate materials.
  
 In this work, we present the theoretical analysis of the dynamic structure factor (DSF) and the linewidth of the CR due to ionized impurity scattering of the 2D electron systems in quantizing magnetic fields. Electrons are assumed to form a Wigner crystal in which the electrons are localized and oscillate around their equilibrium positions, and the electron correlation is taken into account through the Wigner phonons. The numerical results of the CR linewidth are compared with experiments in the Al$_x$Ga$_{1-x}$As heterostructures
  
 The paper is organized as follows. In Sec.~\ref{sec:rela}, the basic relations for the conductivity, linewidth and DSF are presented.  In Sec.~\ref{sec:dyn}, the expression and the results of numerical calculation of the DSF are given. We then evaluate the CR linewidth in Sec.~\ref{sec:cr}. A comparision with the experiment for density denpendence of the CR linewidth is also given in this section. The double peak structure similar to the case of the pinning of charge density waves\cite{FL78} of the diagonal conductivity is shown in Sec.~\ref{sec:cond}. Finally, we summarize our results and give concluding remarks in Sec.~\ref{sec:sum}.

 \section{\label{sec:rela}Basic relations} 

 The absorption lineshape of the 2D electron system in a magnetic field is characterized by the conductivity $\sigma_\pm(\omega)=\sigma_{xx}(\omega)\pm i\sigma_{xy}(\omega)$ which may be expressed through the memory function $M(\omega)$ by:
 \begin{equation}
\sigma_\pm(\omega)=\frac{in_ee^2}{m^*}\frac{1}{\omega\mp\omega_c+M(\omega)},
\end{equation}
where $\omega_c\!=\!eB/m^*c$ is the cyclotron frequency, $-e, m^*, n_e$ are the charge, effective mass and 2D density of the electrons, respectively.

We assume that the 2D electrons forms a Wigner crystal and the 2D position vector of the $n$th-electron is given by ${\bf r}_{n}={\bf R}_{n}+{\bf u}_{n}$, where ${\bf R}_{n}$ is the lattice site vector and ${\bf u}_{n}$ is the deviation from ${\bf R}_{n}$. Further the scattering of electrons is caused by the static impurity potential $V({\bf r})$ which is long-ranged:
\begin{equation}
V({\bf r})=\sum_{\{{\bf R}_i\}}{\frac{e^2}{\kappa[({\bf r}-{\bf R}_i)^2+z_i^2]^{1/2}}},
\end{equation}
where (${\bf R}_i,z_i$) is the coordinate of the $i$th-impurity which is distributed randomly, and $\kappa$ the dielectric constant of the substrate matter. In this case, the density-density correlation function $S({\bf q},t)$ is defined as\begin{equation}\label{eq:srr}
S({\bf q},t)=\sum_{\bf R}{e^{i{\bf q\cdot R}}<e^{-i{\bf q\cdot u}_n(t)}e^{-i{\bf q\cdot u}_m(0)}>},
\end{equation}
where ${\bf R}={\bf R}_n-{\bf R}_m$ and $<\cdots>$ denotes the thermal average.

The memory function $M(\omega)$ is related to the density-density correlation function by 
\begin{eqnarray}
\label{eq:memo}M(\omega)\!\!\!&=&\!\!\!i\frac{1-e^{-\hbar\omega\beta}}{\omega}\frac{1}{A}\sum_{\bf q}{U_q\frac{\hbar q^2}{4m^*}S(q,\omega)}\\
\!\!\!\!&&\!\!\!\!+\frac{1}{A}\!\sum_{\bf q}{U_q\frac{\hbar q^2}{4m^*}\!\!\displaystyle\int\limits_{0}^{\beta}{\!\!\frac{e^{-\hbar\omega(\beta-\tau)}-1}{\omega}S(q,-i\tau)d\tau}},\nonumber
\end{eqnarray}
where $\beta$ is the inverse temperature, $A$ the area of the electron system, and $S({\bf q},\omega)$ is the Fourier transform of $S({\bf q},t)$
\begin{equation} 
S(q,\omega)=\displaystyle\int\limits_{-\infty}^{+\infty}{dt e^{i\omega t}S({\bf q},t)}\label{eq:sr},
\end{equation}
and will be called the dynamic structure factor (DSF) hereafter. In eq.~(\ref{eq:memo}), $U_q$ is the Fourier transform of the random averaged correlation function of ionized impurity potentials $U({\bf r})~\!=~\!<~\!\!V({\bf r})V(0)>_r$ given by\cite{Saitoh98}
\begin{equation}
U_q=\left(\frac{\pi e^2}{\kappa}\right)^2\frac{n_i}{q^2}{\rm erfc}(d/\Delta).
\end{equation}
Here, $n_i$ denotes the impurity concentration and the impurity distribution is characterized by $<z_i>_r=-d$ and $<(z_i+d)^2>_r=\Delta^2/2$.
 
 The imaginary part of the memory function $M(\omega)$ will be called the width function and denoted by $\gamma(\omega)$ since it characteries the absorption line shape. We have:
\begin{equation}
\gamma(\omega)\equiv{\rm Im} M(\omega)=\frac{1-e^{-\hbar\omega\beta}}{\omega}\frac{1}{A}\sum_{\bf q}{U_q\frac{\hbar q^2}{4m^*}{\rm Re}S(q,\omega)}.
\end{equation}

 In order to evaluate this quantity, the calculation of the DSF (\ref{eq:sr}) will be given in the next section.

\section{\label{sec:dyn}Dynamic structure factor}
\vskip 0.3cm
\subsection{Wigner phonon spectra}
\vskip 0.3cm
In the absence of the magnetic field and impurities, Wigner phonon frequencies in the long-wavelength limit  are approximately given by\cite{Bon77} 
 \begin{eqnarray}
 \omega^{(0)}_{{\bf k}l}&\approx&\omega_0\sqrt{\frac{k}{k_0}},\ \ \ {(\rm longitudinal)}\\
 \omega^{(0)}_{{\bf k}t}&\approx&\frac{\sqrt{2}}{4}\omega_0\frac{k}{k_0},\ \ \ {(\rm transverse)}
 \end{eqnarray}
 where $\omega_0=2(\pi n_e)^{3/4}(e^2/\kappa m^*)^{1/2}$ and $k_0=\sqrt{4\pi n_e}$.

In the presence of ionized impurities, the Wigner phonon spectra may have a gap \cite{Saitoh98}
\begin{equation}\label{eq:omega}
\omega_{{\bf k}\lambda}^2=\omega_{{\bf k}\lambda}^{(0)2}+v^2,
\end{equation}
where gap $v$ is given by
\begin{equation}
v^2\approx\frac{1}{A}\sum_{\bf q}{U_q\frac{\mid {\bf q\cdot \epsilon}_\lambda({\bf k})\mid^2}{m^*}\int_0^\beta{S({\bf q},-i\tau)d\tau}}
\end{equation} 
in the longwavelength limit. Here $\epsilon_\lambda({\bf k})$ is the polarization vector of the Wigner phonons with wave vector ${\bf k}$ and mode $\lambda$ in the absence of magnetic fields:
\begin{equation}
\epsilon_l({\bf k})=\frac{\bf k}{k},\ \ \ \epsilon_t={\bf e}_z\times\epsilon_l({\bf k}),
\end{equation}
where ${\bf e}_z$ is the unit vector normal to the electron plane.

When a magnetic field is applied normal to the electron plane, it is well-known that the longitudinal and transverse modes of the Wigner phonons are coupled and split into two modified modes with frequencies $\Omega_{{\bf k}j} (j=\pm)$ given by 
\begin{equation}
\Omega_{{\bf k}\pm}=\frac{\sqrt{\omega_c^2+(\omega_{{\bf k}l}+\omega_{{\bf k}t})^2}\pm \sqrt{\omega_c^2+(\omega_{{\bf k}l}-\omega_{{\bf k}t})^2}}{2}.
\end{equation}

The Fourier transform of the displacement vector ${\bf u}_n$ and the real momentum ${\bf \Pi}_n$ are given, respectively, by
\begin{eqnarray}
\label{eq:u}{\bf u}({\bf k})\!\!\!\!&=&\!\!\!\!\sum_{j=\pm}{\sqrt{\frac{\hbar}{2m^*\Omega_{{\bf k}j}}}\left[{\bf E}_j({\bf k})b_{{\bf k}j}+{\bf E}^*_j({\bf k})b^\dagger_{-{\bf k}j}\right]},\\
{\bf\Pi}({\bf k})\!\!\!&\equiv&\!\!\!{\bf p}({\bf k})+\frac{m^*\omega_c}{2}{\bf e}_z\times{\bf u}({\bf k})\nonumber\\
\!\!\!\!\!\!&=&\!\!\!\!-i\sum_{j=\pm}\!\!{\sqrt{\frac{m^*\hbar\Omega_{{\bf k}j}}{2}}\!\!\left[{\bf E}_j({\bf k})b_{{\bf k}j}\!-{\bf E}^*_j({\bf k})b^\dagger_{-{\bf k}j}\!\right]},
\end{eqnarray}
where ${\bf p}({\bf k})$ is the Fourier transform of the canonical momentum, $b_{{\bf k}j}$ and $b^\dagger_{{\bf k}j}$ are the annihilation and creation operators of the phonon, respectively, and

\begin{equation}
{\bf E}_\pm({\bf k})=\sqrt{\frac{\!\Omega_{{\bf k}\pm}^2-\omega_{{\bf k}t}^2}{\Omega_{{\bf k}+}^2- \Omega_{{\bf k}-}^2}}{\bf e}_l({\bf k})\mp i\sqrt{\frac{\!\Omega_{{\bf k}\pm}^2-\omega_{{\bf k}l}^2}{\Omega_{{\bf k}+}^2- \Omega_{{\bf k}-}^2}}{\bf e}_t({\bf k})\label{eq:pola}.\end{equation}

\subsection{Dynamic structure factor}
\vskip 0.5cm

In order to obtain the expression of the DSF, we calculate the density-density correlation function defined by eq.~(\ref{eq:srr}). We assume here that the correlation between two electrons at the same site is much larger than that of electrons at different sites  and disregard the contributions with ${\bf R}\neq {\bf 0}$ in $S({\bf q},t)$ (single-site approximation). Using eq.~(\ref{eq:u}) we get
\begin{eqnarray}
S({\bf q},t)\!\!\!&\approx&\!\!\!<e^{-i{\bf q\cdot u}_n(t)}e^{i{\bf q\cdot u}_n(0)}>\nonumber\\
\!\!\!\!\!\!&=&\!\!\!\exp{[W({\bf q},t)-W({\bf q},0)]},
\end{eqnarray}
where \begin{equation}
W({\bf q},t)=\sum_{j=\pm}{W_j({\bf q},t)},
\end{equation}
\begin{eqnarray}
W_j({\bf q},t)\!\!\!&=&\!\!\!\frac{\hbar}{2m^*N}\sum_{\bf k}{\frac{\mid{\bf q\cdot E}_j({\bf k})\mid^2}{\Omega_{{\bf k}j}}}\nonumber\\
\!\!\!\!\!\!&\times&\!\!\!\!\left[(1+n_{{\bf k}j})e^{-i\Omega_{{\bf k}j}t}+n_{{\bf k}j}e^{i\Omega_{{\bf k}j}t}\right],
\end{eqnarray}
where $N$ is the number of the lattice sites and $n_{{\bf k}j}$ the Bose-Einstein distribution function of the $({\bf k}j)$-phonon.

When the magnetic field is very high such that $\beta\hbar\omega_c\gg~\!1$ and $\omega_c\gg\omega_0$, we have $\Omega_{{\bf k}+}^2~\!\!\approx~\!\!\omega_c^2+v^2\approx\omega_c^2$ and $\Omega_{{\bf k}-}^2\approx\omega_{{\bf k}t}^2\omega_{{\bf k}l}^2/\omega_c^2\ll \beta^{-2}$, and therefore
\begin{eqnarray}
W_+({\bf q},0)-W_+({\bf q},t)\!\!\!&=&\!\!\!x(1-e^{-i\omega_ct}),\\
W_-({\bf q},0)-W_-({\bf q},t)\!\!\!&=&\!\!\!\frac{x\Gamma^2}{4\hbar^2}(t^2-i\hbar\beta t),
\end{eqnarray}
where $x=\hbar q^2/2m^*\omega_c$ and $\Gamma$ is defined by
\begin{equation}
\Gamma^2=\frac{2\hbar}{\beta\omega_c}\frac{1}{N}\sum_{{\bf k}\lambda}{\omega^2_{{\bf k}\lambda}},
\end{equation}
and will be called the broadening parameter of the DSF.

From eq.~(\ref{eq:omega}), $\Gamma^2$ can be expressed as the sum of two contributions from the electron-electron interaction and from the electron-impurity scattering:
\begin{equation}
\Gamma^2=\Gamma_{\rm e-e}^2+\Gamma_{\rm e-i}^2,
\end{equation}
where $\Gamma_{\rm e-e}$ is related to the electron-electron interaction:
\begin{equation}
\Gamma^2_{\rm e-e}=\frac{2\hbar}{\beta\omega_c}\frac{1}{N}\sum_{{\bf k},\lambda}{\omega^{(0)2}_{{\bf k}\lambda}}=\frac{(0.85\hbar\omega_0)^2}{\beta\hbar\omega_c} 
\end{equation}
and $\Gamma_{\rm e-i}$ describes the single-electron effect related to the electron-impurity scattering:
\begin{equation}
\Gamma^2_{\rm e-i}=\frac{4\hbar v^2}{\beta\omega_c}=\frac{2\hbar}{\pi\beta\tau_i},
\end{equation}
with $\tau_i$ being the collision time due to the impurity scattering.

After performing integration with respect to ${\bf k}$ and $t$, we obtain the final expression of the DSF \cite{An2003} 

\begin{eqnarray}\label{eq:sqf}
S(q,\omega)\!\!\!&=&\!\!\!\frac{2\sqrt{\pi}\hbar}{\Gamma}\sum_{n=0}^{\infty}{\frac{x^{n-\frac{1}{2}}}{n!} e^{\displaystyle\beta\hbar(\omega-n\omega_c)/2}}\nonumber\\
\!\!\!\!\!\!\!\!\!\!&\times&\!\!\!\!\!\!\!\exp{\!\left[\!-x\left(\!1\!+\!\left(\!\frac{\Gamma\beta}{4}\right)^2\!\right)\!-\frac{\hbar^2(\omega-n\omega_c)^2}{x\Gamma^2}\!\right]}\!.
\end{eqnarray} 

It is interesting to see that the structure of the DSF (\ref{eq:sqf}) is quite similar to the DSF of the highly correlated electron liquid by Monarkha {\it et al}. \cite{Monarkha2000}. However, it differs from ours in that the expression of $\Gamma_c$ in ref.~9 corresponds to our expressions by $\Gamma_c~=~x^{-1}\sqrt{\Gamma_{\rm e-i}^2+x\Gamma_{\rm e-e}^2}$. Monarkha {\it et al}. introduced the correlation effect phenomenologically by assuming the Doppler shift due to the ultra-fast electrons. Moreover, the factor $\exp{[\beta\hbar(\omega-n\omega_c)/2-x(\Gamma\beta/4)^2]}$ is missing in their results, leading to an overestimation of the contributions from the higher Landau levels $n>1$. 
  \begin{figure}[h]
\begin{center}\leavevmode 
\includegraphics[width=0.9\linewidth,height=0.8\linewidth]{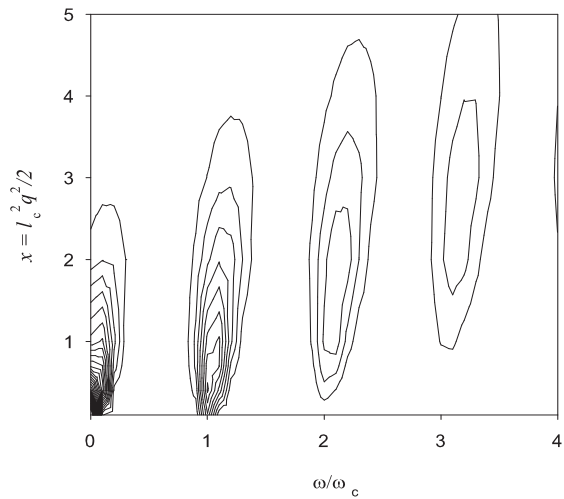}
\caption{\label{fig:dsf} Contour map of $S(q,\omega)$ using material constants of GaAs/AlGaAs at $B=7.4$~T, $T=4.2$~K and $n_e~\!=~\!2\cdot~10^{10}$~cm$^{-2}$. Impurity concentration is set to be $n_i=0.1n_e$ \cite{Chui94}, and $d/\Delta=1.70$.}\end{center}\end{figure}
  
   In the limit $q\to 0$, the DSF (\ref{eq:sqf}) has a singularity $S(q,\omega)\propto \delta(\omega)$ which is consistent with Kohn theorem \cite{Kohn61}. At $q\neq 0$, the DSF has a series of peaks at points ($\omega\approx n\omega_c+x\Gamma^2\beta/4\hbar, x\approx n-1/2$) as shown in Fig.~\ref{fig:dsf} and the peak widths are broadened not only by the electron interaction but also by the impurity scattering through the parameter $\Gamma$. The peak height deceases with increasing $x$ and becomes very small at $x\gg 1$. Hence the most importance contributions into DSF come from the terms with $n\leq 1$. 
 
  In the extremely high magnetic fields and $\omega=\omega_c$ (cy~\!\!clotron resonance regime) when $\Gamma$ can be considered very small, all electrons at the lowest Landau level, and thus the term with $n=1$ dominates the DSF. In this case, our theory reproducts the results obtained previously by Dyk~\!\!man \cite{Dykman82}. Otherwise, the contributions of higher Lan~\!\!dau levels must be taken into account.
  
  \section{\label{sec:cr}Cyclotron resonance linewidth}
  
  Using eq.~(\ref{eq:sqf}), we rewrite the CR linewidth of the 2D correlated electrons under quantizing magnetic fields in more explicit form
  \begin{eqnarray}
  \gamma(\omega_c)\!\!\!&=&\!\!\!\displaystyle\frac{\Gamma_{\rm e-i}^2}{\hbar\Gamma}\sinh{\frac{\beta\hbar\omega_c}{2}}\nonumber\\
  \!\!\!\!\!\!&\times&\!\!\!\!\!\sum_{n=0}^{\infty}{\frac{e^{-n\beta\hbar\omega_c/2}}{n!(2\xi^2)^{n+\frac{1}{2}}}F_n\left(\frac{2\xi\hbar\omega_c\mid 1-n\mid}{\Gamma}\right)},
  \end{eqnarray}  
  where $\xi=\sqrt{1+(\Gamma\beta/4)^2}$ and the function $F_n(z)$ is defined by  \begin{eqnarray}
\displaystyle F_n(z)=\left\{
	\begin{array}{@{\,}ll}
\displaystyle\frac{\pi}{\sqrt{2}}, &\hbox{(for $n=1$)},\\
 \displaystyle{\frac{2}{\sqrt{\pi}}}\displaystyle z^{n+1/2}K_{n+\frac{1}{2}}(z), &\hbox{(for $n\neq 1$)}.
	\end{array}
	\right.
\end{eqnarray}
Here, $K_\nu(z)$ stands for the modified Bessel function of the second kind.
\begin{figure}[h]
\begin{center}\leavevmode
\includegraphics[width=1.0\linewidth,height=0.92\linewidth]{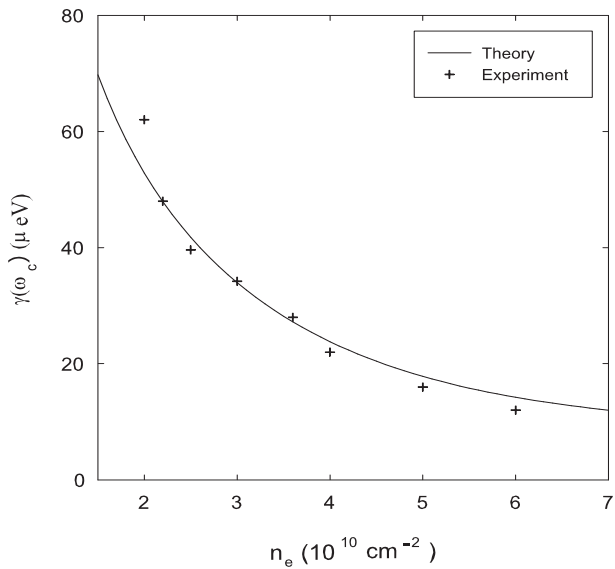}
\caption{\label{fig:ne1} $\gamma(\omega_c)$ {\it vs}.\!~electron density for the 2D electrons in heterostructure GaAs/AlGaAs at $T\!=\!4.2$~K, $B\!=\!7.4$~T. The crosshairs are taken from the experimental results of Chou {\it et al}.\cite{Chou88}.}\end{center}\end{figure}
 
 The electron density dependence of the CR linewidth is shown in Fig.~\ref{fig:ne1}. The CR linewidth monotonously decreases with increasing electron density. Our theoretical result is in quantitative agreement with the experiment in the region of the electron density $2~\!\cdot~\!10^{10}$~cm$^{-2}~\le~n_e\le~6\cdot~10^{10}$~cm$^{-2}$ (the filling factor $\nu=2\pi n_e(c\hbar/eB)$ varies from 0.08  to 0.34), where the electron system is expected to crystalize into a Wigner crystal.
\begin{figure}[h]
\begin{center}\leavevmode
\includegraphics[width=1.0\linewidth,height=0.81\linewidth]{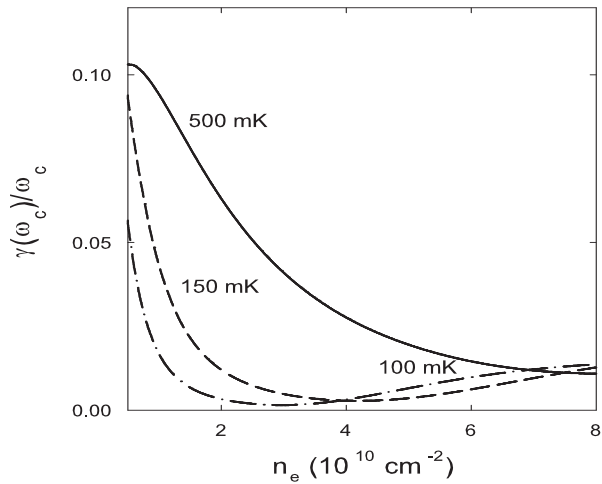}
\caption{\label{fig:ne2}$\gamma(\omega_c)/\omega_c$ {\it vs.} $n_e$ at $B=10$~T and for several temperatures $T= 500$~mK (solid curve), $150$~mK (dashed curve), and $100$~mK (dotted curve).}\end{center}\end{figure}

 Figure~\ref{fig:ne2} illustrates the density dependence of the CR linewidth at lower temperatures. It is easy to see that the density denpendence of the CR linewidth is sensitive to the change of the temperature. At relatively high temperatures ($T\ge 500$~mK), the CR linewidth decreases monotonically with $n_e$. For lower temperatures ($T\le 150$~mK), the CR linewidth has a shallow minimum around  $n_e\approx 3\cdot~\!10^{10}$~cm$^{-2}$ and then increases slightly. At sufficiently low temperatures ($T< 150$~mK), where $\Gamma_{\rm e-e}$ can be considered smaller than $\Gamma_{\rm e-i}$ due to impurities, so that $\Gamma\approx \Gamma_{\rm e-i}$, the role of impurities becomes most important.
\begin{figure}[h]
\begin{center}
\includegraphics[width=1.0\linewidth,height=0.81\linewidth]{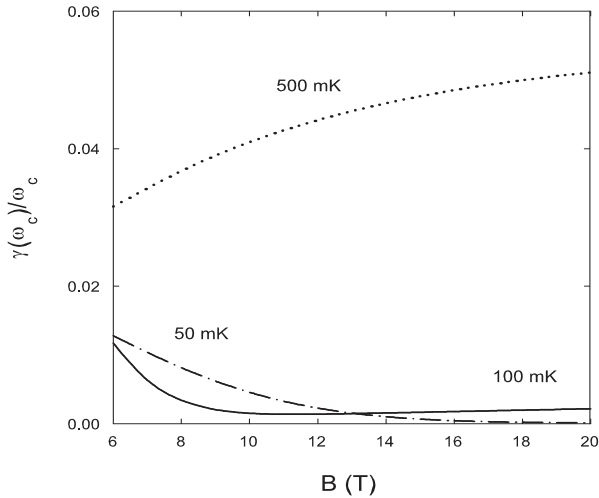}
\caption{\label{fig:B}$\gamma(\omega_c)/\omega_c$ {\it vs}. magnetic field at $n_e=3~\!\!\cdot~\!\!10^{10}$~cm$^{-2}$ for several temperatures: 500~mK (dotted cuve), 200~mK (dashed curver), 100~mK (solid curve) and 50~mK (dash-dotted curve).}\end{center}\end{figure}

Figure~\ref{fig:B} shows the magnetic field dependence of the ratio $\gamma(\omega_c)/\omega_c$ at $n_e=3\cdot 10^{10}$~cm$^{-2}$ for different temperatures. The different behavours in the magnetic field dependence of the CR linewidth at several values of temperatures are clearly seen in this figure. At sufficiently high temperatures ($T>200$~mK), $\gamma(\omega_c)/\omega_c$ increases monotonically with $B$. At $T=200$~mK, $\gamma(\omega_c)/\omega_c$ first decreases, and then slightly increases with $B$.  At $T~\!=~\!100$~mK, $\gamma(\omega_c)/\omega_c$ becomes nearly constant for large $B$ ($B>9$~T). This means that CR linewidth is propotional to $B$ at $T\approx 100$~mK. For $T<100$~mK, $\gamma(\omega_c)/\omega_c$ decreases for a whole range of $B$.

\newpage
\section{\label{sec:cond}Double peak structure of the conductivity}
Next, we turn to the diagonal conductivity. It is determined from eq.~(1) by
\begin{equation}
\sigma(\omega)=\frac{1}{2}(\sigma_+(\omega)+\sigma_-(\omega)).
\end{equation}
\begin{figure}[h] 
\begin{center}\leavevmode
\includegraphics[width=0.90\linewidth,height=0.9\linewidth]{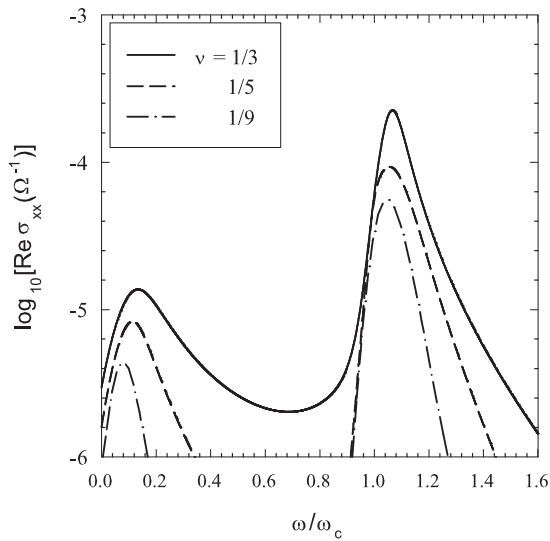}
\caption{\label{fig:conduc}Frequency dependence of the diagonal conductivity for se~\!\!veral filling factors at $B=6.15$~T and $T=1$~K.}\end{center}\end{figure}
Figure~\ref{fig:conduc} exhibits the real part of the diagonal conductivity Re$\sigma_{xx}(\omega)$ as a function of $\omega/\omega_c$ for several Landau filling factors $\nu$. Two  peaks are visible in the conductivity. It is well-known that Wigner phonon spectra in a magnetic field split into two modified modes $\Omega_\pm$ as given by eq.~(13). In a high magnetic field case, from eq.~(13) we have $\Omega_+\approx \omega_c$ and $\Omega_-\approx v^2/\omega_c\approx (2\pi\tau_i)^{-1}$.  The peak at $\omega=\omega_c$ corresponds to the ordinary CR, and one at lower frequency $\omega=\Omega_-$ is related to the pinning mode due to impurities. This behavour is similar to the case of the charge density wave pinned due to impurities in the presence of magnetic fields. Fukuyama and Lee developed the CR theory for the charge density waves with the short-ranged pinning potential \cite{FL78}. The magnetic field was treated semiclassically and their result for $\Omega_-$ is depends on the bulk compressibility of the material. In our case, the pinning is determined by the ionized impurities which is long-ranged, and the pinning mode $\Omega_-$ is independent of material constants of the semiconductor substrate where electrons are located.

\section{\label{sec:sum}Conclusion}
 
 We have presented in this paper a theoretical analysis of the cyclotron resonance of the 2D electron system in heterostructures under quantizing magnetic fields at finite temperature. We studied the DSF and the CR linewidth when the electron correlation effect is taken into account through Wigner phonons. We have expressed  the DSF as a summation over all Landau levels and obtained the analystic expression of the CR linewidth. By taking the electron correlation into account through Wigner phonons, the single-electron and many-electron effects are  considered simultaneously not phenomenologically as was done in previous theories\cite{Monarkha2000}. The electron density dependence of the CR linewidth is shown to be in the quantitative agreement with experimental results by Chou et al\cite{Chou88} in the regime where the crystallization of electrons is expected. 
 
 Similar to the case of the charge density waves with the short-ranged pinning potential \cite{FL78}, our results of the conductivity also show the doubling of the resonance peaks. But unlike the charge density wave case, the pinning mode in our theory is independent of the bulk compressibility of the substrate materials.
 
 We have studied also other range of parameters than experiments. The numerical calculations  shows that the CR linewidth monotonously decreases with increasing electron density at sufficiently high temperatures. On the other hand, the density dependence at low temperatures ($T<150$ mK) is quite different: a sharp decrease in the very low density region ($n_e\le 2~\!\cdot~\!10^{10}$~cm$^{-2}$) and  a slight increase for higher densities. The dependence of the CR linewidth on the magnetic field is also evaluated. Specifically, at sufficiently high temperatures ($T>100$~mK), the cyclotron resonance is broadened considerably when the magnetic field becomes high, and at lower temperatures the broadening of the CR peak becomes very small.

 \vskip 0.4cm
\noindent{\bf Acknowledgments}
 \vskip 0.3cm
 One of authors (V. A. Dinh) acknowledges the financial support from the Ministry of Education and Science, Sports and Culture of Japan.

\end{document}